\documentclass[twocolumn]{aastex631}

\usepackage{xcolor}
\usepackage{threeparttable}

\def\Oiii{[O\,{\sc iii}]}

\def\micron{$\mu$m}
\def\lsun{$L_{\rm \odot}$}
\def\msun{$M_{\rm \odot}$}
\def\ltsima{$\buildrel<\over\sim$}
\def\la{\lower.5ex\hbox{\ltsima}~}
\def\gtsima{$\buildrel>\over\sim$}
\def\ga{\lower.5ex\hbox{\gtsima}~}
\def\deg~{$^{\circ}$}

\shorttitle{AASTeX v6.3.1 Sample article}
\shortauthors{Hashimoto et al.}
\graphicspath{{./}{figures/}}
\begin{document}

\title{Reionization and the ISM/Stellar Origins with JWST and ALMA (RIOJA):
The core of the highest redshift galaxy overdensity at $z = 7.88$ confirmed by NIRSpec/JWST}

\correspondingauthor{T. Hashimoto}
\email{hashimoto.takuya.ga@u.tsukuba.ac.jp}
\author[0000-0002-0898-4038]{T. Hashimoto}
\affiliation{Division of Physics, Faculty of Pure and Applied Sciences, University of Tsukuba,Tsukuba, Ibaraki 305-8571, Japan}
\affiliation{Tomonaga Center for the History of the Universe (TCHoU), Faculty of Pure and Applied Sciences, University of Tsukuba, Tsukuba, Ibaraki 305-8571, Japan}
\author[0000-0002-7093-1877]{J. \'Alvarez-M\'arquez}
\affiliation{Centro de Astrobiolog\'{\i}a (CAB), CSIC-INTA, Ctra. de Ajalvir km 4, Torrej\'on de Ardoz, E-28850, Madrid, Spain}
\author[0000-0001-7440-8832]{Y. Fudamoto}
\affiliation{Waseda Research Institute for Science and Engineering, Faculty of Science and Engineering, Waseda University, 3-4-1 Okubo, Shinjuku, Tokyo 169-8555, Japan}
\affiliation{National Astronomical Observatory of Japan, 2-21-1, Osawa, Mitaka, Tokyo, Japan}
\author[0000-0002-9090-4227]{L. Colina}
\affiliation{Centro de Astrobiolog\'{\i}a (CAB), CSIC-INTA, Ctra. de Ajalvir km 4, Torrej\'on de Ardoz, E-28850, Madrid, Spain}
\author[0000-0002-7779-8677]{A. K. Inoue}
\affiliation{Waseda Research Institute for Science and Engineering, Faculty of Science and Engineering, Waseda University, 3-4-1 Okubo, Shinjuku, Tokyo 169-8555, Japan}
\affiliation{Department of Physics, School of Advanced Science and Engineering, Faculty of Science and Engineering, Waseda University, 3-4-1, Okubo, Shinjuku, Tokyo 169-8555, Japan}
\author[0000-0002-0984-7713]{Y. Nakazato}
\affiliation{Department of Physics, The University of Tokyo, 7-3-1 Hongo, Bunkyo, Tokyo 113-0033, Japan}
\author[0000-0002-8680-248X]{D. Ceverino}
\affiliation{Universidad Autonoma de Madrid, Ciudad Universitaria de Cantoblanco, E-28049 Madrid, Spain}
\affiliation{CIAFF, Facultad de Ciencias, Universidad Autonoma de Madrid, E-28049 Madrid, Spain}
\author[0000-0001-7925-238X]{N. Yoshida}
\affiliation{Department of Physics, The University of Tokyo, 7-3-1 Hongo, Bunkyo, Tokyo 113-0033, Japan}
\affiliation{Kavli Institute for the Physics and Mathematics of the Universe (WPI), UT Institute for Advanced Study, The University of Tokyo, Kashiwa, Chiba 277-8583, Japan}
\affiliation{Research Center for the Early Universe, School of Science, The University of Tokyo, 7-3-1 Hongo, Bunkyo, Tokyo 113-0033, Japan}
\author[0000-0001-6820-0015]{L. Costantin}
\affiliation{Centro de Astrobiolog\'{\i}a (CAB), CSIC-INTA, Ctra. de Ajalvir km 4, Torrej\'on de Ardoz, E-28850, Madrid, Spain}
\author[0000-0001-6958-7856]{Y. Sugahara}
\affiliation{Waseda Research Institute for Science and Engineering, Faculty of Science and Engineering, Waseda University, 3-4-1 Okubo, Shinjuku, Tokyo 169-8555, Japan}
\affiliation{National Astronomical Observatory of Japan, 2-21-1, Osawa, Mitaka, Tokyo, Japan}
\author[0000-0003-2119-277X]{A. Crespo G\'omez}
\affiliation{Centro de Astrobiolog\'{\i}a (CAB), CSIC-INTA, Ctra. de Ajalvir km 4, Torrej\'on de Ardoz, E-28850, Madrid, Spain}
\author[0009-0005-5448-5239]{C. Blanco-Prieto}
\affiliation{Centro de Astrobiolog\'{\i}a (CAB), CSIC-INTA, Ctra. de Ajalvir km 4, Torrej\'on de Ardoz, E-28850, Madrid, Spain}
\author[0000-0003-4985-0201]{K. Mawatari}
\affiliation{Division of Physics, Faculty of Pure and Applied Sciences, University of Tsukuba,Tsukuba, Ibaraki 305-8571, Japan}
\affiliation{Tomonaga Center for the History of the Universe (TCHoU), Faculty of Pure and Applied Sciences, University of Tsukuba, Tsukuba, Ibaraki 305-8571, Japan}
\author[0000-0001-7997-1640]{S. Arribas}
\affiliation{Centro de Astrobiolog\'{\i}a (CAB), CSIC-INTA, Ctra. de Ajalvir km 4, Torrej\'on de Ardoz, E-28850, Madrid, Spain}
\author[0000-0001-8442-1846]{R. Marques-Chaves}
\affiliation{Geneva Observatory, Department of Astronomy, University of Geneva, Chemin Pegasi 51, CH-1290 Versoix, Switzerland}
\author[0000-0002-4005-9619]{M. Pereira-Santaella}
\affiliation{Instituto de F\'isica Fundamental (IFF), CSIC, Serrano 123, E-28006, Madrid, Spain}
\author[0000-0002-5268-2221]{T. J. L. C. Bakx}
\affiliation{Department of Physics, Graduate School of Science, Nagoya University, Nagoya 464-8602, Japan}
\affiliation{National Astronomical Observatory of Japan, 2-21-1, Osawa, Mitaka, Tokyo, Japan}
\author[0000-0001-8083-5814]{M. Hagimoto}
\affiliation{Department of Physics, Graduate School of Science, Nagoya University, Nagoya 464-8602, Japan}
\author{T. Hashigaya}
\affiliation{Department of Astronomy, Kyoto University Sakyo-ku, Kyoto 606-8502, Japan}
\author[0000-0003-3278-2484]{H. Matsuo}
\affiliation{National Astronomical Observatory of Japan,
2-21-1 Osawa, Mitaka, Tokyo 181-8588, Japan}
\affiliation{Graduate University for Advanced Studies (SOKENDAI), 2-21-1 Osawa, Mitaka, Tokyo 181-8588, Japan}
\author[0000-0003-4807-8117]{Y. Tamura}
\affiliation{Department of Physics, Graduate School of Science, Nagoya University, Nagoya 464-8602, Japan}
\author{M. Usui}
\affiliation{Division of Physics, Faculty of Pure and Applied Sciences, University of Tsukuba,Tsukuba, Ibaraki 305-8571, Japan}
\author{Y. W. Ren}
\affiliation{Department of Pure and Applied Physics, Graduate School of Advanced Science and Engineering, Faculty of Science and Engineering, Waseda University, 3-4-1 Okubo, Shinjuku, Tokyo 169-8555, Japan}

\received{2023 May 19}\accepted{2023 August 31}\published{2023 September 14}

\begin{abstract}
The protoclusters in the epoch of reionization, traced by galaxies overdensity regions, are ideal laboratories for studying the process of stellar assembly and cosmic reionization. We present the spectroscopic confirmation of the core of the most distant protocluster at $z = 7.88$, A2744-z7p9OD, with the {\it James Webb Space Telescope} NIRSpec integral field unit spectroscopy. The core region includes as many as 4 galaxies detected in \Oiii\ 4960 \AA\ and 5008 \AA\ in a small area of $\sim 3\arcsec \times 3\arcsec$, corresponding to $\sim$ 11 kpc $\times$ 11 kpc, after the lensing magnification correction. Three member galaxies are also tentatively detected in dust continuum in Atacama Large Millimeter/submillimeter Array Band 6, which is consistent with their red ultraviolet continuum slopes, $\beta \sim -1.3$. The member galaxies have stellar masses in the range of log($M_{*}/M_{\rm \odot}$) $\sim 7.6-9.2$ and star formation rates of $\sim 3-50$ $M_{\rm \odot}$ yr$^{-1}$, showing a diversity in their properties. FirstLight cosmological simulations reproduce the physical properties of the member galaxies including the stellar mass, \Oiii\ luminosity, and dust-to-stellar mass ratio, and predict that the member galaxies are on the verge of merging in a few to several tens Myr to become a large galaxy with $M_{\rm *}\sim 6\times10^{9} M_{\rm \odot}$. The presence of a multiple merger and evolved galaxies in the core region of A2744-z7p9OD indicates that environmental effects are already at work 650 Myr after the Big Bang.
\end{abstract}

\section{Introduction}
\label{sec:intro}

Understanding the properties of galaxies at redshift of $z\gtrsim6$ is crucial for studying galaxy formation and evolution as well as the process of cosmic reioinization. With the advent of the James Webb Space Telescope ({\it JWST}; \citealt{Gardner2023}), it has become possible to study detailed properties of high redshift galaxies (e.g., \citealt{Curtis-Lake2023, Bunker2023}). 

The overdense regions of galaxies, protoclusters, offer us the opportunity to study environmental effects of galaxy formation (e.g., \citealt{Dressler1980, Koyama2008}). Protoclusters in the epoch of reionzation (EoR) are interesting because they probe early galaxy formation and stellar mass assembly.

\begin{figure*}
    \centering
    \includegraphics[height=5.5cm]{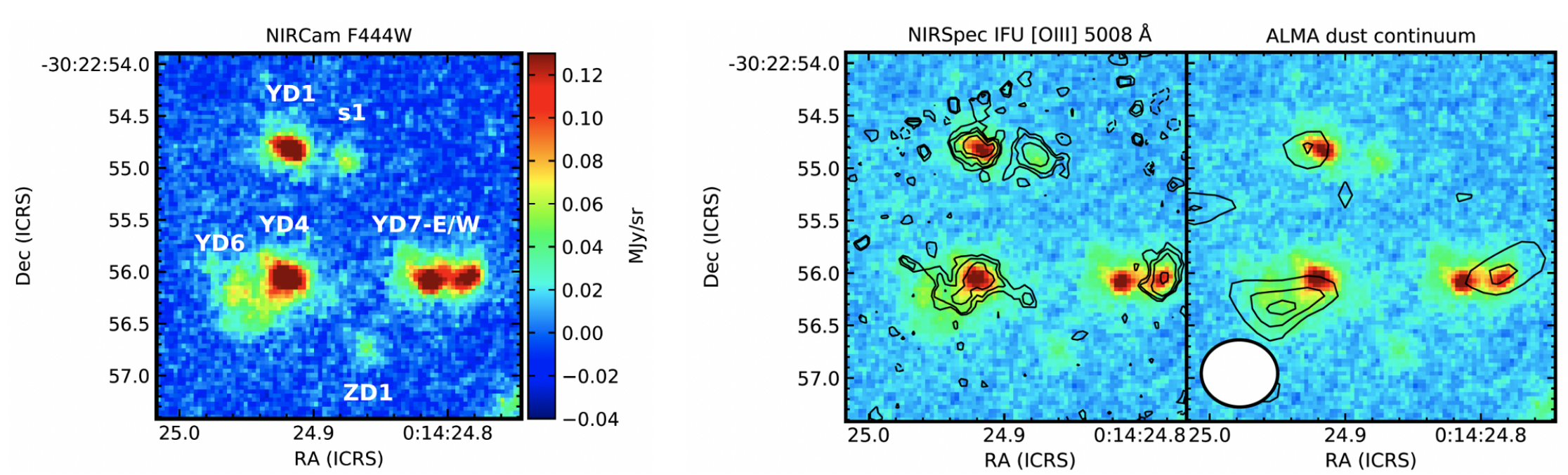}
    \caption{{\bf (Left)} $3\farcs0\times3\farcs0$ cutout image of NIRCam F444W around the ``quintet'' region. 
    The colorbar shows the surface brightness. {\bf (Middle)} Black contours show the NIRSpec \Oiii\ 5008 \AA\ line map overlaid on the NIRCam F444W image. Contours are drawn at $\pm (2, 3, 4, 5, 6)\times \sigma$, where $\sigma$ is 1.47 MJy sr$^{-1}$. {\bf (Right)} Black contours show the ALMA Band 6 dust continuum map at the rest-frame wavelength $\sim 133$ \micron\ overlaid on the F444W image. Contours are drawn at $\pm (2.5, 3.0, 3.3)\times \sigma$, where $\sigma$ is $6.7$ $\mu$Jy beam$^{-1}$. The ellipse at the lower left corner indicates the synthesized beam size ($0\farcs73\times0\farcs64$).}
    \label{fig:summary}
\end{figure*}

An interesting candidate protocluster, A2744-z7p9OD, at $z\approx7-8$ behind the strong lensing cluster Abell 2744 was reported based on photometric redshift deduced from {\it HST} data taken as a part of Hubble Frontier Fields Survey (PI: Lotz) in conjunction with {\it Spitzer}/IRAC data (\citealt{Zheng2014, Laporte2014, Ishigaki2016}). 
The protocluster includes eleven member galaxy candidates at $7 < z < 8$ labelled as A2744-ZD1, ZD2, ..., ZD11 and eleven candidates at $8 < z  < 9$ labelled as A2744-YD1, YD2, ..., YD11 within a single field of view of {\it HST} WFC3. Among these, particularly interesting is the ``quintet'' galaxies at $z\sim8$ (YD1, YD4, YD6, YD7, and ZD1) in a distance of $\sim 3\arcsec \times3\arcsec$, as shown in the left panel of Figure \ref{fig:summary}. 

Recently, \cite{Morishita2022} have reported the spectroscopic redshift of A2744-z7p9OD by identifying seven galaxies at $z=7.88$ with e.g., \Oiii\ 4960.295 \AA\ and 5008.239 \AA\ (hereafter 4960 \AA\ and 5008 \AA), identifying the system as  the most distant protocluster reported so far. Observations were made by JWST ERS (program \#1324, PI: Treu) and DDT (program \#2756, PI: Chen) programs using NIRSpec (\citealt{Jakobsen2022}) micro shutter array (MSA; \citealt{Ferruit2022}) mode with a prism ($R\sim100$) and/or medium-resolution ($R\sim1000$) configurations. 

The combination of JWST NIRSpec integral field spectroscopy (IFS; \citealt{Boker2022}) and Atacama Large Millimeter/submillimeter Array (ALMA) is a promising route to identify  member galaxies in a less biased manner, and study their properties such as dust content and star formation history in a spatially resolved manner. As part of the JWST GO1 program \#1840 (PIs: J. \'{A}lvarez-M\'{a}rquez and T. Hashimoto; \citealt{GO1840}), the Reionization and the ISM/Stellar Origins with JWST and ALMA (RIOJA) project, we are conducting follow-up observations of twelve ALMA \Oiii\ 88 \micron\ emission line galaxies at $z \sim 6 - 8$ (e.g., \citealt{Inoue2016, Laporte2017, Carniani2017, Hashimoto2019, Harikane2020, Akins2022, Witstok2022, Sugahara2022}) with NIRCam (\citealt{Rieke2005, Beichman2012}) and NIRSpec IFS. 
Here we report observations of the ``quintet'' galaxy candidates at $z\sim8$ (Fig. \ref{fig:summary}). Two of the member galaxies (A2744-YD4 and YD7) were already identified in \cite{Morishita2022} at $z=7.88$. YD4 was also detected in dust continuum emission (\citealt{Laporte2017}). 
Throughout this letter, we assume a set of parameters in $\Lambda$CDM cosmology with $\Omega_{\small m} = 0.272$, $\Omega_{\small b} = 0.045$, $\Omega_{\small \Lambda} = 0.728$ and $H_{\small 0} = 70.4$ km s$^{-1}$ Mpc$^{-1}$ (\citealt{Komatsu2011}). We adopt the Kroupa initial mass function \citep{Kroupa02}. All fluxes are given without correction by magnification factor of $\mu=2.0$ (\citealt{Morishita2022}) unless otherwise indicated.

\section{JWST Observations of the ``quintet''}
\label{sec:jwst}
\subsection{NIRSpec IFS Observations}
\label{sec:nirspec_IFS}

NIRSpec IFS observations were conducted on 16 Dec 2022 as part of the JWST GO1 program \#1840. The specific observations can be accessed via the Mikulski Archive for Space Telescopes (MAST) at the Space Telescope Science Institute: \dataset[doi:10.17909/bn0s-2879]{https://doi.org/10.17909/bn0s-2879}. The observations were taken with a grating/filter pair of G395H/F290LP that produced a cube with a spectral resolution of $R\sim2700$ in a wavelength range of $2.87-5.27$ \micron. A medium size cycling dither pattern of four dithers was used to cover all the quintet galaxies at slightly larger distances than the $3\farcs0\times3\farcs0$ IFS field-of-view. The total exposure time including overhead was 8870 seconds. 

The raw data were processed with the JWST pipeline version 1.11.1 (\citealt{Bushouse2022}) under CRDS context jwst$\_$1097.pmap. Following the data reduction process used by Guaranteed Time Observations programs (e.g., \citealt{Marshall2023, Ubler2023, Perna2023}), we applied some modifications to the pipeline including (1) the removal of the $1/f$ noise (c.f., \citealt{Bagley2023}), (2) rejection of the bad pixels and cosmic rays by sigma-clipping, and (3) removal of a median background in the calibrated images. The data cube was sampled with a pixel size of $0\farcs05$. To set the astrometry in the NIRCam reference system, we cross-correlated the \Oiii\ 5008 \AA\ map with the NIRCam F444W image, the latter significantly contaminated by the \Oiii\ 5008 \AA\ emission.

\subsection{NIRCam Observations }
\label{subsec:imaging}

NIRCam data were taken as a part of the JWST GO1 program \#2561, the UNCOVER survey (PI: Labbe \& Bezanson; \citealt{UNCOVER}). The data include seven filters, F115W, F150W, F200W, F277W, F356W, F410M, and F444W. 
The data were calibrated with the JWST pipeline version 1.9.4 under CRDS context jwst$\_$1041.pmap. Apart from the standard pipeline stages, we applied custom snowball and wisps correction as described in \citet{Bagley2023}. Furthermore, following \citet{PerezGonzalez2023}, a background homogenization algorithm was applied prior to obtaining the final mosaics, all drizzled at a plate scale of $0\farcs03$ px$^{-1}$. The angular resolution is $\approx 0\farcs14$ in F444W (Fig. \ref{fig:summary}). 

We rescaled the photometric errors to take into account the correlated noise introduced by drizzling in the NIRCam images. We used 1000 circular apertures ($r=0\farcs225$) on blank regions to measured the rms on both the nominal and drizzled images (see also \citealt{PerezGonzalez2023}, L. Costantin et al.~in preparation). This effect was estimated to be $\sim$60\% for LW bands and 3\% on SW bands.

\subsection{Identification of galaxies at $z=7.88$}

\begin{figure}
    \centering
    \vspace{-1cm}
    \includegraphics[width=7cm]{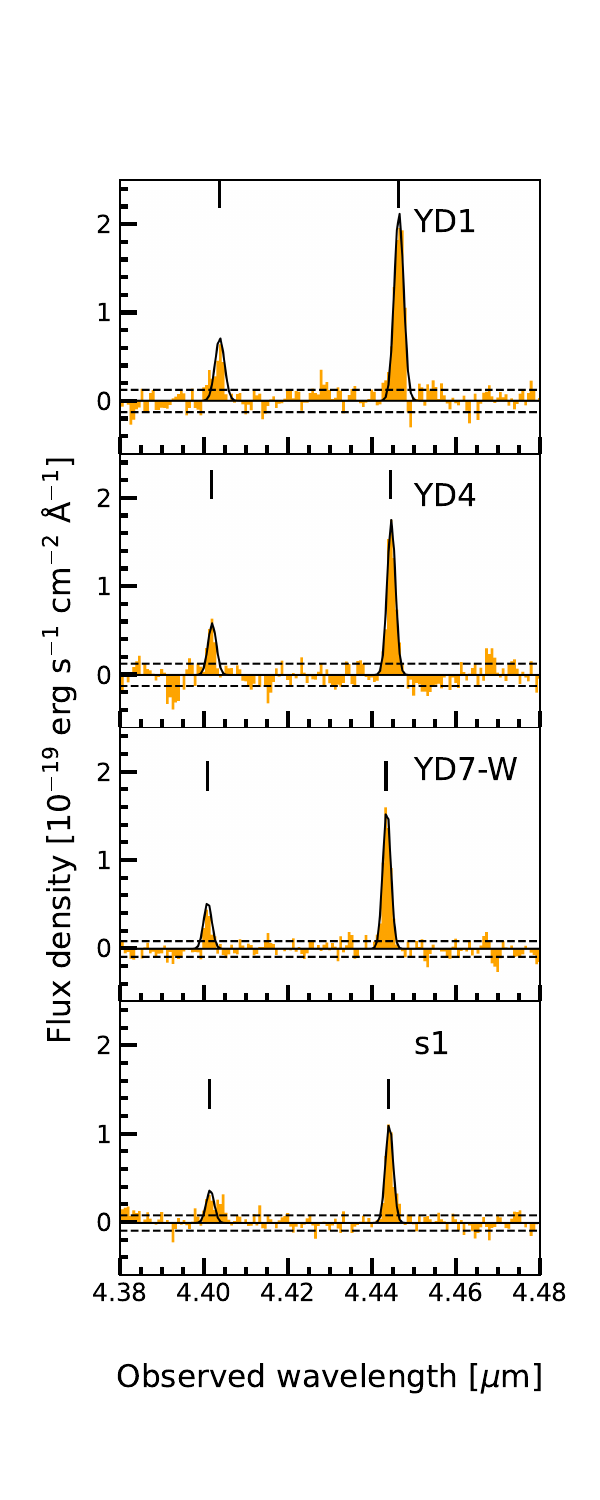}
    \vspace{-1cm}
    \caption{Spectra of \Oiii\ 4960\AA\ and 5008 \AA\ associated with A2744-YD1, YD4, the western component of YD7 (YD7-W), and s1. 
    Black solid lines show Gaussian fit results for emission lines, whereas the horizontal dashed lines indicate the typical $1\sigma$ uncertainties in the spectra measured around the \Oiii\ 5008 \AA\ lines. The vertical ticks indicate the wavelengths of \Oiii\ 4960 and 5008 \AA.} 
    \label{fig:spectra}
\end{figure}

The map of the \Oiii\ 5008 \AA\ line emission integrated over a velocity range of 270 km s$^{-1}$ is overlaid on the NIRCam F444W image in Figure \ref{fig:summary} (central panel). The NIRSpec spectra covering the \Oiii\ 4960 \AA\ and 5008 \AA\ spectral range for the four galaxies with detected \Oiii\ emission are presented in Figure \ref{fig:spectra}. The spectra were spatially integrated over the region corresponding to a $2\sigma$ region in the \Oiii\ 5008 \AA\ map, where $1\sigma$ is 1.47 MJy sr$^{-1}$. 

In addition to the three known member galaxies, A2744-YD1, YD4, and the western component of YD7 (YD7-W), we serendipitously identified a new galaxy, A2744-s1, that was not recognized as a $z\sim8$ candidate in the previous works based on the HST and Spitzer data. This demonstrates the power of IFS observations to probe the true structure of a protocluster. A2744-ZD1 was not identified in our spectra, which indicates its faint nature in \Oiii.

Table \ref{tab:SEDresults} summarizes the line properties. We measured the spectroscopic redshift, line FWHM, and flux by applying a Gaussian fit to the \Oiii\ 5008 \AA\ lines alone. To obtain uncertainties, we first measured the noise level per spectral bin at $\lambda = 4.2-4.3$ and $4.6-4.8$ \micron, and adopted its standard deviation as the $1\sigma$ value. The $1\sigma$ value was then considered in the Gaussian fitting. We obtained the spectroscopic redshifts ranging from $7.8721$ to $7.8778$, corresponding to the maximum velocity offset of $193\pm10$ km s$^{-1}$.
The line FWHM values are measured to be $110-145$ km s$^{-1}$ after correction by the instrumental broadening (FWHM$_{\rm inst} =$ 100 km s$^{-1}$; \citealt{Jakobsen2022}). In A2744-ZD1, we place a $3\sigma$ upper limit on the \Oiii\ 5008 \AA\ flux based on a spectrum extracted from a circular aperture with a radius of $0\farcs2$. In this calculation, we assumed the $z=7.88$ and line FWHM of 150 km s$^{-1}$. 
In Fig. \ref{fig:spectra}, we also show the Gaussian curves to the \Oiii\ 4960 \AA\ lines, where we adopted the same line FWHMs and redshifts as \Oiii\ 5008 \AA, and the peak flux density ratio of 3:1 (\citealt{Storey2000}).

A2744-YD4 and YD7 were previously observed in \cite{Morishita2022} with the MSA mode. The flux values in \cite{Morishita2022} and those in the present IFS observations are broadly consistent with each other if we take the slit-loss correction factors into account (private communication with T. Morishita).

\section{ALMA Dust Continuum Data of A2744-z7p9OD}
\label{subsec:data3}

The spectroscopic identification of the galaxies motivates us to examine dust continuum with ALMA data. The dust continuum was already detected in A2744-YD4 at the significance level of $4\sigma$ in ALMA Band 7 data (\citealt{Laporte2017}). In Appendix, we re-analyzed the Band 7 data, and present the photometry for the member galaxies. Below, we additionally examine  ALMA archival data in Band 6. 

The Band 6 data were obtained as a part of an ALMA Cycle 6 program (ID: 2018.1.01332.S, PI: N. Laporte). A single tuning with four spectral windows was used to probe a mean frequency of 254.2 GHz, corresponding to the rest-frame wavelength of 132.8 \micron\ at $z=7.88$. 

We reduced and analyzed ALMA data with the default CASA pipeline scripts. We produced a continuum map using all the four spectral windows with a natural weighting to optimize the point-source sensitivity. We adopted the taper value of $0\farcs4$ where the signal-to-noise ratio becomes the highest. The beam size of the tapered map is  $0\farcs73\times0\farcs64$, and the rms value is 6.7 $\mu$Jy beam$^{-1}$. 

The right panel of Figure \ref{fig:summary} shows the dust continuum contours overlaid on the F444W image. The dust continuum contours show a spatial coincidence in A2744-YD1, YD4, and YD7, within positional uncertainties, at a significance level $>3\sigma$. 
The continuum flux densities are $83\pm34$, $60\pm25$, and $53\pm25$ $\mu$Jy in A2744-YD1, YD4, and YD7, respectively, without lensing magnification correction. 
We note that the spatial resolution of the ALMA Band 6 data is insufficient to spatially resolve A2744-YD4 at $z=7.88$ and YD6 that was reported to be $z=8.38$ (\citealt{Laporte2017}) with a separation of only $0\farcs45$ in the sky coordinate. 
In Appendix, we analysed the positional uncertainty of the Band 6 continuum emission along with the previous Band 7 continuum data of \cite{Laporte2017}. These analyses demonstrate that there is a possibility that a part of the Band 6 continuum emission in YD4 could be contaminated by YD6, whereas the Band 7 continuum is likely associated with YD4. 
In A2744-ZD1 and s1, we place a $3\sigma$ upper limit on the flux density to be $20$ $\mu$Jy. 

We estimate the total infrared luminosity, $L_{\rm IR}$, by integrating the modified black-body radiation over $8-1000$ \micron\footnote{Only in the case of YD4, we use the Band 7 photometry, not Band 6 photometry, due to the potential contamination by YD6 (see Appendix).}. Because dust temperature and emissivity index values cannot be constrained with the current data, we assume a dust temperature of $T_{\rm d}=$ 50 K and an emissivity index of $\beta_{\rm d}=1.5$ (e.g., \citealt{Bakx2021}). We obtain $L_{\rm IR}\approx $ ($1.4\pm0.5$), ($1.6\pm0.6$), and ($0.8\pm0.4$) $\times 10^{11}$ \lsun\ for A2744-YD1, YD4, and YD7, respectively, after correcting for the cosmic microwave background effects (\citealt{da_Cunha2013}). The dust mass is estimated to be $M_{\rm d}=$ ($2.7\pm1.1$), ($3.3\pm1.2$), and ($1.8\pm0.8$) $\times 10^{6}$ \msun, for A2744-YD1, YD4, and YD7, respectively, with a dust mass absorption coefficient $\kappa = \kappa_{0} (\nu/\nu_{0})^{\beta_{\rm d}}$, where $\kappa_{0} = 10$ cm$^{2}$ g$^{-1}$ at 250 \micron\ (\citealt{hildebrand1983}). Note that these physical quantities are corrected for the magnification factor of 2.

\begin{figure*}
    \centering
    \includegraphics[width=\textwidth]{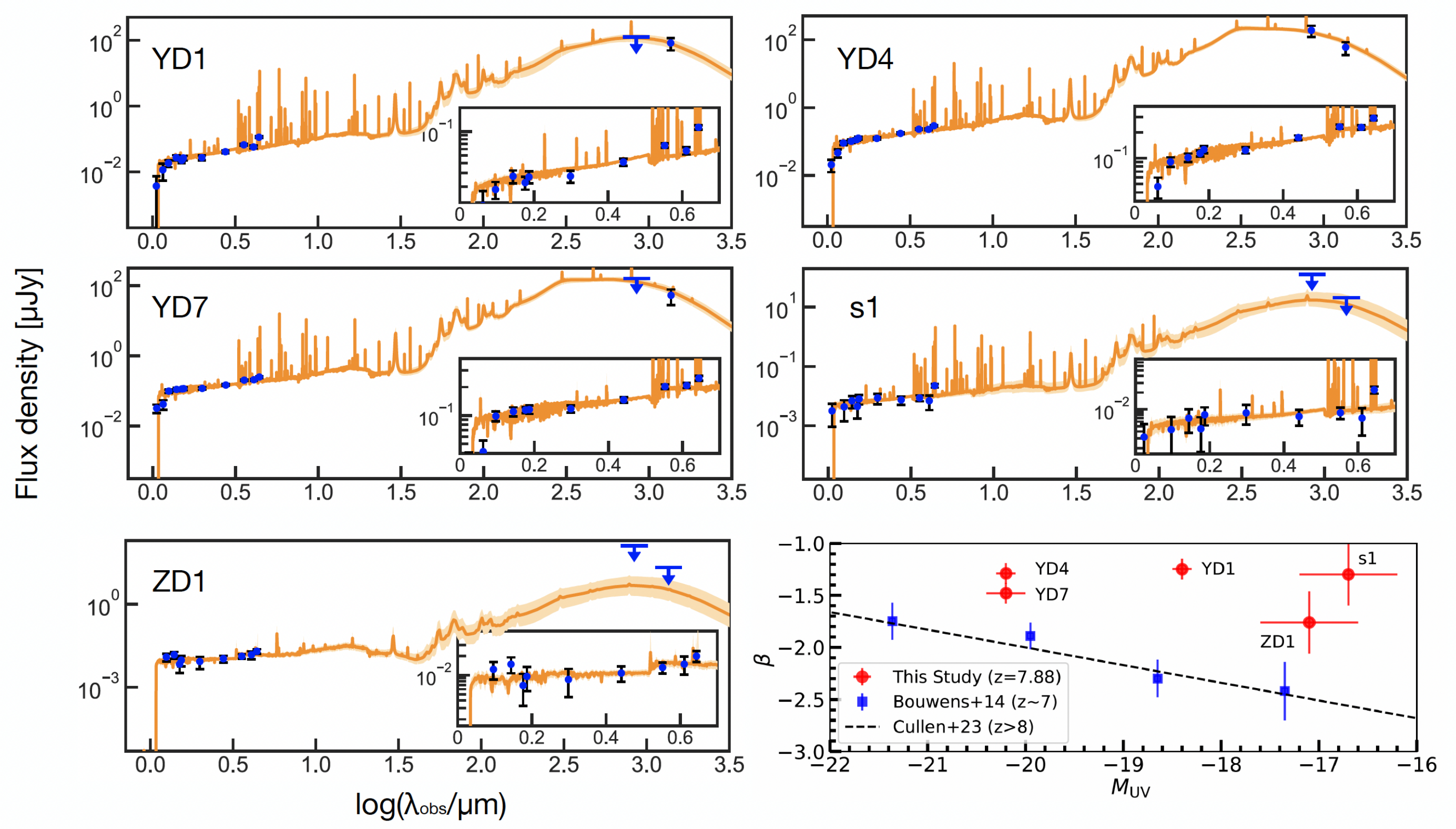}
    \caption{Photometry and results of BAGPIPES SED fitting. Blue points show the aperture photometry using JWST, HST, and ALMA. Downward arrows show $3\,\sigma$ upper limits for ALMA's dust continuum non-detections. Orange solid lines show 50th percentile of the posterior distribution, and the orange bands show 16th to 84th percentile of the posterior distribution. Insets show zoom in of the SEDs for the rest-frame UV and optical wavelengths. {\bf (Bottom right)} The blue squares show the typical observed $\beta$ as a function of $M_{\rm UV}$ at $z\sim7$ (\citealt{Bouwens2014}). The dashed line indicates the best-fit relation between the two quantities at $z >8$ (\citealt{Cullen2023}). The galaxies in the ``quintet'' region show redder $\beta$ (red circles).}
    \label{fig:fullSED}
\end{figure*}

\section{SED fitting}

To obtain physical properties of the galaxies, we performed SED fitting using photometry obtained from JWST and ALMA observations.
We also use the ancillary HST data from the UNCOVER project (\citealt{Weaver2023}) that include six filters, F435W, F606W, F814W, F125W, F140W, and F160W. 
Details of our methods are presented in Y. Fudamoto et al.~in preparation, and in the following we summarize essential items.

We used the publicly available SED fitting code \texttt{BAGPIPES} \citep{Carnall2018} assuming delayed-tau star formation history (SFH).  We adopt a dust attenuation law from \citet{Calzetti00}. For A2744-ZD1, we assume the redshift of $z=7.88$. 
The \Oiii\ line and dust continuum fluxes are included in the  fitting process, where we correct for the CMB impact on the dust continuum flux densities following \citet{da_Cunha2013} and assuming $T_{\rm d} = 50$ K. For A2744-s1 and ZD1 in which dust continuum is not detected, the SED fitting uses only the HST and JWST photometry. Then, we checked if the obtained SEDs are consistent with the non-detections of dust continuum. Likewise, for A2744-YD1 and YD7 in which only Band 6 data show the detection, we first run the SED fitting without Band 7 data, and checked if the results are consistent with the non-detections in Band 7.

We obtained reasonable fits as shown in Figure \ref{fig:fullSED}. Table \ref{tab:SEDresults} summarizes the physical properties. We  estimated UV continuum slopes ($\beta$), using full SED posteriors by fitting a power-law function in a form of $f_{\lambda}\propto\lambda^{\beta}$. The member galaxies have red UV continuum slopes ranging from $\beta = $ $-1.2$ to $-1.8$, and high stellar dust attenuation ranging from $A_{\rm V} =$ 0.2 to 0.8 magnitudes, supporting that the member galaxies are relatively dusty and can be classified as luminous infrared galaxies. The present sample has a median $A_{\rm V}$/mag value of 0.64 with a standard deviation of 0.20, which is large compared to that of 13 photometrically selected galaxies at $z\sim7-9$, $A_{\rm V}$/mag $=0.20\pm0.18$ (\citealt{Leethochawalit2023}).

The galaxies have stellar masses in the range of log($M_{*}/M_{\rm \odot}$) $\sim 7.6-9.2$ and star formation rates of $\sim 3-50$ $M_{\rm \odot}$ yr$^{-1}$, showing a wide variety in stellar properties. We note that the SFR values become twice lower if we adopt a constant SFH.

\begin{figure*}
    \centering
    \includegraphics[scale=0.75]{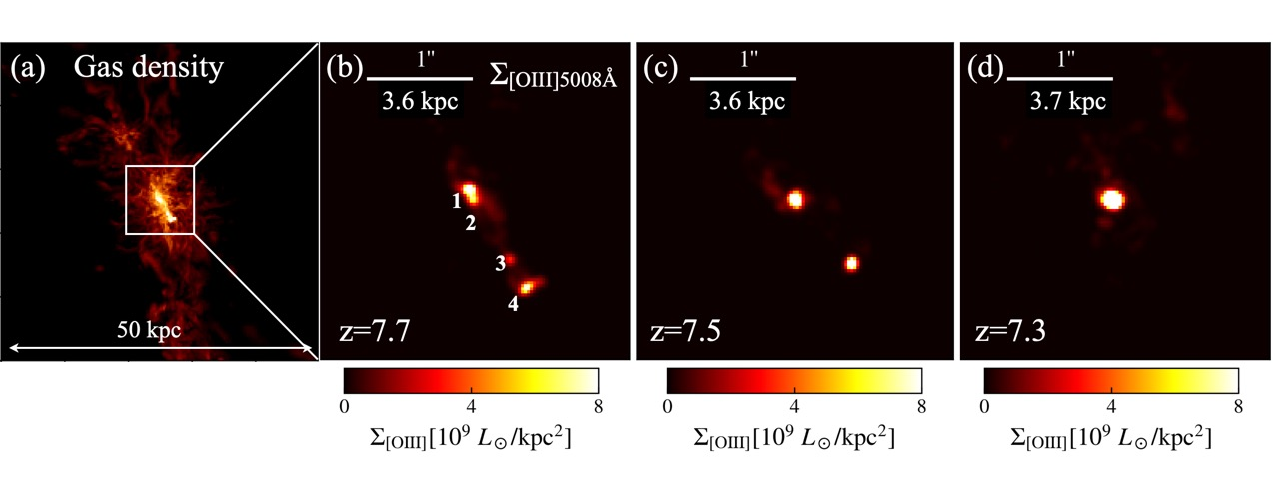}
    \caption{Projected gas density (panel a) and \Oiii$\,5008\,{\rm \AA}$  distribution (panel b) for a galaxy sample FL957 (\citealt{Ceverino2017}) at $z = 7.7$. Each panel shows a region with a side length and depth of 50 kpc and 10.6 kpc, the latter corresponds to the NIRSpec IFS field of $3\arcsec$ with magnification of $\mu \sim 2$. Panels (c) and (d) show the following evolution of FL957 at $z=7.5$ and $7.3$, demonstrating that the system finally becomes a large massive galaxy by $z=7.3$.}
    \label{fig:FL957_projection}
\end{figure*}

\begin{deluxetable*}{cccccccccc}[t] 
\tablecaption{Summary of Properties of the ``quintet'' galaxies\label{tab:SEDresults}}
\tablewidth{0pt}
\tablehead{
\colhead{ID} & \colhead{$z$}  & \colhead{$f_{\rm [OIII]5008}$} & \colhead{FWHM$_{\rm [OIII]5008}$} & \colhead{$\beta$} & \colhead{$M_{\rm UV}$} & \colhead{A$_{\rm V}$}& \colhead{SFR$_{\rm 10 Myr}$} & \colhead{log$M_{\ast}$} & \colhead{log$M_{\rm d}$}\\ 
\colhead{} & \colhead{} & \colhead{$10^{-18}$ erg s$^{-1}$ cm$^{-2}$} & \colhead{km s$^{-1}$} & \colhead{} & \colhead{mag} & \colhead{mag}& \colhead{$M_{\odot}\,yr^{-1}$}& \colhead{$M_{\odot}$}& \colhead{$M_{\odot}$}
}
\startdata
YD1 & $7.8778\pm0.0001$ & $6.00\pm0.38$ & $145\pm9$ & $-1.25^{+0.11}_{-0.14}$ & $-18.4\pm0.1$ & $0.84^{+0.20}_{-0.25}$ & $15^{+4}_{-3}$ & $8.4^{+0.1}_{-0.1}$ & $6.4$\\ 
YD4 & $7.8742\pm0.0001$ & $4.34\pm0.36$ & $122\pm10$ & $-1.29^{+0.08}_{-0.10}$ & $-20.2\pm0.1$ & $0.65^{+0.13}_{-0.08}$ & $45^{+10}_{-7}$  & $9.2^{+0.1}_{-0.1}$ & $6.5$\\
YD7 & $7.8721\pm0.0001$ & $3.78\pm0.25$ & $113\pm7$ & $-1.48^{+0.09}_{-0.10}$ & $-20.2\pm0.2$ & $0.50^{+0.15}_{-0.08}$ & $31^{+8}_{-5}$ & $9.1^{+0.1}_{-0.1}$ & $6.3$\\ 
s1 & $7.8732\pm 0.0001$ & $2.63\pm0.24$ & $110\pm10$ & $-1.30^{+0.32}_{-0.32}$ & $-16.7\pm0.5$ & $0.64^{+0.25}_{-0.20}$ & $3^{+2}_{-1}$ & $7.6^{+0.3}_{-0.3}$ & $<5.9$\\ 
ZD1 & $7.88^{\dagger}$ & $< 0.26 $ & N/A & $-1.76^{+0.30}_{-0.23}$ & $-17.1\pm0.5$ & $0.24^{+0.16}_{-0.11}$ & $1.0^{+0.6}_{-0.3}$ & $8.1^{+0.1}_{-0.1}$ & $<5.9$ 
\enddata
\tablecomments{
The SED fitting results and dust masses are corrected for the gravitational magnification factor of 2 as measured in  \citet{Morishita2022}. We calculated SFR from the SFH posterior distributions by averaging star formation rates over $10\,{\rm Myr}$. Upper limits are $3\sigma$.
}
\tablenotetext{\dagger}{
We assumed redshift of $z=7.88$ for A2744-ZD1 during SED fitting. 
}

\end{deluxetable*}

\section{Discussion}
\label{sec:discussion}

\subsection{A2744-z7p9OD. An early $z\sim8$ protocluster dust factory}
The ``quintet'' region demonstrates the accelerated evolution of galaxies. The UV continuum slope is a useful tracer of dust content (e.g., \citealt{Meurer1999}), and has been  systematically investigated in e.g., \cite{Bouwens2014} and \cite{Cullen2023} at $z\sim4-8$ and $z\sim8-16$, respectively. As shown in the bottom right panel of Figure \ref{fig:fullSED}, these studies have shown that UV fainter galaxies have smaller $\beta$ (i.e., bluer) values, indicating that they are less dusty. The red circles indicate the member galaxies, where the UV absolute magnitude, $M_{\rm UV}$, ranges from $-20.2$ to $-16.7$ after the lensing magnification correction (Table \ref{tab:SEDresults}). The member galaxies show redder $\beta$ at a given $M_{\rm UV}$, indicating that they are dustier. 

Albeit with large uncertainties in the dust mass estimates due to unknown dust temperature, we roughly estimate the dust-to-stellar mass ratio, log($M_{\rm d}/M_{\rm *}$), to be $\approx$ $-2.0$, $-2.7$, and $-2.8$ in A2744-YD1, YD4, and YD7, respectively. These  are consistent with the median value obtained in a sample of local spiral galaxies (KINGFISH; \citealt{Kennicutt2011}), log($M_{\rm d}/M_{\rm *}$) $=-3.03^{+0.51}_{-0.78}$ (\citealt{Calura2017}). The mass ratio is also consistent with the result of semi-analytical galaxy formation models of DELPHI (\citealt{Dayal2022}) that takes into account the Type II supernovae (SNe) dust production, astration, shock destruction, and ejection in outflows. 

At $z\sim8$, a dominant fraction of dust is produced by Type II SNe. Based on the stellar mass estimate and an effective number of SNe per unit stellar mass, $0.15\times10^{-2}$ $M_{\rm \odot}^{-1}$ (\citealt{inoue2011}), we estimate the number of SNe events to be $\approx (4-20)\times10^{6}$ for the three galaxies. To reproduce the dust mass, we estimate the dust yield per unit SN to be $0.7-1.0$ $M_{\rm \odot}$, which can be reproduced by theoretical models in the case of effective  production and inefficient destruction of dust grains (e.g., \citealt{Lesniewska2019}). Note that the dust yield is reduced if these galaxies have (i) a top-heavy IMF and/or (ii) dust temperature higher than the assumed value of 50 K.

\subsection{A2744-z7p9OD. Progenitor protocluster and evolution according FirstLight simulations}

It is interesting to see if an evolved system as the ``quintet region'' is also found in cosmological simulations. To this end, we use FirstLight simulations \citep{Ceverino2017}, which are a suite of ‘zoom-in’ cosmological simulations with a parsec-scale resolution and
provide a larger number of galaxy samples among
other simlations
\citep[e.g.,][]{Pallottini2017a, Katz2019, Arata2020}. The cosmological boxsize is 60 comoving Mpc and the maximum spatial resolution is $\sim9-17$ pc, which are well-suited to study the internal structure of rare objects such as the ``quintet''. 

We have successfully found an analogue of the ``quintet'' region. Panel (a) of Figure \ref{fig:FL957_projection} shows the density projection of the simulation sample, FL957, at $z=7.7$. There are four galaxies within a $3 \arcsec \times 3\arcsec$ region, and there is yet another stellar clump within $\sim$ $1\farcs5$ ($\sim$ 5 kpc). The stellar masses of the four clustered galaxies are $M_{*} =$ 4.2, 5.8, 8.6, and 7.2 $\times 10^{8}$ \msun, consistent with the observed values within a factor of 2 (Table \ref{tab:SEDresults}). The maximum velocity offset among the four simulated clumps is 131 km s$^{-1}$, similar to the observed value of $193\pm10$ km s$^{-1}$.
We calculate the \Oiii$\,5008\,{\rm \AA}$ luminosity in the same manner as in \citet{Nakazato2023}. Panel (b) of Figure \ref{fig:FL957_projection} shows the \Oiii$\,5008\,{\rm \AA}$ line surface brightness, where we assume dust extinction of $A_{5008}=$ 0.25 mag following the relation between stellar mass and attenuation as reported in \citet{Mushtaq2023} for FirstLight simulation. 
The total line luminosity in the region is $1.95\times 10^{43} ~\mathrm{erg~s^{-1}}$, which is consistent with the total intrinsic luminosity of the observed galaxies, $0.62\times 10^{43}~\mathrm{erg~s^{-1}}$, within a factor of 3. Finally, assuming a fixed dust-to-metal ratio of 0.4, the dust-to-stellar mass ratio is log$(M_{\rm d}/M_{\rm *})$ $\sim -2.5$, again consistent with the observed values.

The simulations allow us to follow the further evolution of the system. We find that galaxy 2 and 3 merge at $z=7.5$ (panel c), and finally become a massive galaxy with $M_{\rm *} = 5.47 \times 10^9~M_\odot$ at $z=7.3$ (panel d). The rapid evolution implies that the ``quintet'' galaxies are likely on the verge of merging in a few to several tens Myrs. Detailed star formation history and gas dynamics will be studied in our forthcoming paper (Y. Nakazato et al.~in preparation).

\section{Conclusions}

We have presented the JWST NIRSpec IFS observations of the core of the most distant protocluster at $z=7.88$, A2744-z7p9OD. 
\begin{itemize} 
\item We have identified four galaxies at $z=7.88$ with \Oiii\ 4960 \AA\ and 5008 \AA\ within a $\sim$ 11 kpc $\times$ 11 kpc region. The galaxies have a small velocity offset of $193\pm10$ km s$^{-1}$, suggesting a close asscociation of these galaxies. 
\item ALMA Band 6 dust continuum data show $3\sigma$ signals at the positions of A2744-YD1, YD4+YD6, and YD7, whereas Band 7 data show $4\sigma$ signal at the position of YD4. Assuming a dust temperature of 50 K and an emissivity index of 1.5, these three galaxies have $L_{\rm IR} \sim 10^{11} L_{\rm \odot}$ and $M_{\rm d} \sim (1-3)\times 10^{6} M_{\rm \odot}$. 
\item The presence of dust is consistent with their red UV continuum slopes, $\beta \sim -1.3$. The galaxies have redder $\beta$ values at a given $M_{\rm UV}$ compared to typical $z\gtrsim7$ galaxies. 
\item With the BAGPIPES SED fitting code assuming delayed-tau SFH, the member galaxies have stellar masses in the range of log($M_{*}/M_{\rm \odot}$) $\sim 7.6-9.2$ and star formation rates of $\sim 3-50$ $M_{\rm \odot}$ yr$^{-1}$, showing a diversity in their properties. 
\item Following FirstLight cosmological simulations reproducing the observed physical properties of the member galaxies, the expected evolution of the core protocluster is to form a larger, massive galaxy with a stellar mass of $M_{\rm *}\sim 6\times10^{9} M_{\rm \odot}$ at $z\sim7.3$.
\end{itemize}

The present study demonstrates the power of JWST NIRSpec IFS observations to probe the true structure of a protocluster. The presence of a multiple merger and evolved galaxies in the core region of A2744-z7p9OD indicates that environmental effects are already at work 650 Myr after the Big Bang.

\section*{Acknowledgments}

\begin{acknowledgments}
We thank an anonymous referee for valuable comments that have greatly improved the paper.
This paper makes use of the following ALMA data: ADS/JAO.ALMA \#2015.1.00594S and \#2018.1.01332.S. ALMA is a partnership of ESO (representing its member states), NSF (USA) and NINS (Japan), together with NRC (Canada), NSC and ASIAA (Taiwan), and KASI (Republic of Korea), in cooperation with the Republic of Chile. The Joint ALMA Observatory is operated by ESO, AUI/NRAO and NAOJ.
This work has made use of data from the European Space Agency (ESA) mission {\it Gaia} (\url{https://www.cosmos.esa.int/gaia}), processed by the {\it Gaia} Data Processing and Analysis Consortium (DPAC, \url{https://www.cosmos.esa.int/web/gaia/dpac/consortium}). Funding for the DPAC has been provided by national institutions, in particular the institutions participating in the {\it Gaia} Multilateral Agreement.
TH was supported by Leading Initiative for Excellent Young Researchers, MEXT, Japan (HJH02007) and by JSPS KAKENHI Grant Number 22H01258.  
AKI, YS, and YF are supported by NAOJ ALMA Scientific Research Grant Numbers 2020-16B. AKI also acknowledges funding from JSPS KAKENHI Grant Number 23H00131. 
YT is supported by JSPS KAKENHI Grant Number 22H04939. K.M. acknowledges financial support from the Japan Society for the Promotion of Science (JSPS) through KAKENHI grant No. 20K14516.
J.A-M., L.C., A.C-G., C.B-P acknowledge support by grant PIB2021-127718NB-100 from the Spanish Ministry of Science and Innovation/State Agency of Research MCIN/AEI/10.13039/501100011033 and by "ERDF A way of making Europe". C.B-P acknowledge support by grant CM21\_CAB\_M2\_01 from the Program "Garant\'ia Juven\'il" from the "Comunidad de Madrid" 2021. 
LC acknowledges financial support from Comunidad de Madrid under Atracci\'on de Talento grant 2018-T2/TIC-11612.
MPS acknowledges funding support from the Ram\'on y Cajal program of the Spanish Ministerio de Ciencia e Innovaci\'on (RYC2021-033094-I).
We are grateful to Takahiro Morishita, Nicolas Laporte, Guido Roberts-Borsani, Seiji Fujimoto, and Hide Yajima for useful discussion. 
\end{acknowledgments}

\vspace{5mm}
\facilities{JWST(STIS), ALMA}

\software{astropy \citep{astropy2022}, photoutils \citep{Bradley2022} , BAGPIPE \citep{Carnall2018}, CLOUDY \citep{Ferland2017}
          }

\appendix

\section{ALMA Band 7 data}

We have re-analyzed the archival ALMA Band 7 data (ID: 2015.1.00594S, PI: N. Laporte). The data probe a mean frequency of 356.0 GHz, corresponding to the rest-frame wavelength of 94.8 \micron\ at $z=7.88$. The continuum image has a rms level of 23.4 $\mu$Jy per beam at the phase center of ($\alpha_{\rm ICRS}$, $\delta_{\rm ICRS}$) = ($00^h 14^m 25\fs 081, -30\arcdeg 22' 49\farcs70$), with a beam size of $0\farcs25\times0\farcs18$. Because the core region of A2744-z7p9OD is $\sim 7\arcsec$ away from the phase center, we need to correct for the the primary beam attenuation. With a correction factor of 0.65, the effective rms level of the continuum image is 36.1 $\mu$Jy per beam. 
The continuum image reveals a $4\sigma$ detection at the position of ($\alpha_{\rm ICRS}$, $\delta_{\rm ICRS}$) = ($00^h 14^m 24\fs 939, -30\arcdeg 22' 56\farcs06$). With a positional uncertainty of $\approx 0\farcs06$ (see equation 10.7 of the ALMA Cycle 10 technical handbook), this signal is likely asscociated with YD4, as first reported by \cite{Laporte2017}. The flux density is estimated to be $189\pm68$ $\mu$Jy, which is larger than the previous measurement in \cite{Laporte2017} due to the primary beam attenuation correction.  
Other member galaxies at $z\sim7.88$ (YD1, YD7, ZD1, s1) and YD6 at $z=8.38$ are not detected above $3\sigma$. With the $3\sigma$ upper limit of the brightness, 108 $\mu$Jy per beam, and integrating over the size of these galaxies, the $3\sigma$ upper limits on the flux densities are 127, 127, 160, and 127 $\mu$Jy for A2744-YD1, YD6, YD7, s1, and ZD1, respectively. Note that these values are not corrected for the magnification factor of 2. 

\section{Astrometry of ALMA Band 6 data}

The peak positions of the ALMA Band 6 dust continuum emission for A2744-YD1, YD4+YD6, and YD7 are 
($\alpha_{\rm ICRS}$, $\delta_{\rm ICRS}$) = 
($00^h 14^m 24\fs 930, -30\arcdeg 22' 54\farcs79$), 
($00^h 14^m 24\fs 945, -30\arcdeg 22' 56\farcs35$), 
($00^h 14^m 24\fs 791, -30\arcdeg 22' 55\farcs98$), 
respectively. 
With the peak significance levels of $3.0\sigma$ (YD1), $3.3\sigma$ (YD4+YD6), and $3.1\sigma$ (YD7), and the beam size of the image ($0\farcs73 \times 0\farcs64$), the positional uncertainties of the dust continuum emission are $\approx 0\farcs25$. This indicates that the Band 6 dust continuum around YD4 could be contaminated by YD6.

\section{NIRspec IFS result of A2744-YD6}

\cite{Laporte2017} originally reported $4\sigma$ detections of \Oiii\ 88 \micron\ and Ly$\alpha$ with ALMA and VLT/X-Shooter, respectively, at $z=8.38$ from YD4 (see also \citealt{Carniani2020}). However, the redshift of YD4 is updated to be $z=7.88$ by \cite{Morishita2022}, as reproduced by our IFS observations. The authors discussed the possibility that the source detected at $z=8.38$ is in the background of the protocluster at $z=7.88$, and likely associated with YD6, whereas the dust continuum detection is likely associated with YD4 (see Figure 2 in \citealt{Laporte2019}). 
We examined the presence of an emission line at the position of A2744-YD6; however, our NIRSpec IFS data do not show any optical emission lines such as H$\beta$ or \Oiii\ at $z=8.38$. Assuming a line FWHM of 100 km s$^{-1}$, the $3\sigma$ upper limit for the \Oiii\ 5008 \AA\ line is $0.23\times 10^{-18}$ erg s$^{-1}$ cm$^{-2}$ for YD6.  Future deeper JWST observations are crucial to confirm whether YD6 is at $z=8.38$ using optical emission lines. 


\end{document}